\documentclass{iopart}
\usepackage{graphicx}
\usepackage{iopams}
\usepackage{xcolor}

\begin{document}
\title[Impurity Flow in TJ-II: Incompressibility and Comparison with NC Theory]
{Incompressibility of Impurity Flows in Low Density TJ-II Plasmas and Comparison with Neoclassical Theory}
\author{J Ar\'evalo, J A Alonso, K J McCarthy and J L Velasco}
\address{Laboratorio Nacional de Fusi{\'o}n, Asociaci\'on Euratom-Ciemat, Madrid, Spain}
\ead{juan.arevalo@ciemat.es}
\begin{abstract}
  Poloidal and toroidal velocities of fully-ionised carbon are measured by 
  means of  charge   exchange recombination spectroscopy (CXRS) in 
  the TJ-II stellarator. We present a detailed treatment of the 3D 
  geometry and show that flow measurements, performed at different 
  locations of the same flux surface, are compatible with flow 
  incompressibility for the low density plasmas under study (line 
  averaged electron densities $\bar{n}_e\,\le\,10^{19}$ m$^{-3}$).
  Furthermore, comparison with neoclassical calculations shows 
  quantitative agreement with the measured radial electric field 
  and ion bootstrap parallel flow in the absence of an external momentum input.
  
\end{abstract}
\section{Introduction}
  Plasma rotation has become one of the key ingredients of
  fusion plasma performance, after the discovery of a reduction
  of turbulence and transport~\cite{TerryRevModPhys2000} through a 
  sheared $E\times B$ rotation. With regard to stability, large toroidal rotations 
  can stabilize resistive wall modes and Neoclassical Tearing modes.
  For these reasons, the physics of toroidal 
  momentum transport, and the so-called intrinsic rotation, have become
  a very active area of research, mainly for their importance in the 
  operation of a reactor plasma, for which the external torque will 
  be negligible~\cite{ITER2007chp3}.
  Thus, reliable rotation measurements are requiered. In this 
  sense, the Charge eXchange Recombination Spectroscopy
  (CXRS) technique~\cite{IslerPPCF1994} has developed into 
  a basic tool for diagnosing plasma rotation over the last three decades, 
  and will be a key diagnostic for ITER~\cite{ITER2007chp7}.

  One of the basic results of the kinetic theory of strongly magnetized 
  plasmas with small (compared to thermal) $E\times B$ velocity, is that 
  first order ion and electron flows are parallel to flux surfaces and 
  incompressible~\cite{HintonRMP1976}, \emph{i.e.}, their velocity fields satisfy 
  $\nabla\cdot\mathbf{u} = 0$. This can be understood as a consequence of 
  particle number conservation $\nabla\cdot n\mathbf{u} = 0$ with flux-constant 
  density $n(\rho)$ and zero first-order radial velocity $\mathbf{u}\cdot\nabla \rho = 0$
  (here, $\rho$ is a radial coordinate labelling a magnetic surface).
  Such a velocity field is fully determined by two flux functions 
  (\emph{i.e.}, that depend on spatial coordinates through $\rho$ only). 
  Note that the incompressibility of flows follows from the usual neoclassical 
  (and gyrokinetic) ordering schemes but it neither assumes nor implies 
  the flows to be of 'neoclassical origin'. Indeed, the spatial variation 
  of an incompressible flow has been recently employed to estimate the poloidal 
  velocity from inboard and outboard toroidal measurements in 
  TCV and DIII-D~\cite{CamenemEPS2012,ChrystalRSI2012}. Nevertheless, 
  in the Alcator C-Mod~\cite{MarrPPCF2012} and ASDEX-Upgrade~\cite{PuterichNF2012}
  tokamaks, it has been observed that as the main ion pressure increases, 
  a significant poloidal variation of the impurity density appears. 
  Indeed, a redistribution of the density of impurities on flux surfaces 
  is expected to be driven by the parallel friction caused by large ion diamagnetic 
  velocities~\cite{Helander_PoP1998}, making impurity flows compressible. 

  There is no clear picture concerning the Neoclassical (NC)
  nature of poloidal flows in tokamaks. Early measurements on
  TFTR~\cite{BellPRL1998} showed a large discrepancy 
  with the NC theory. However, a deeper understanding of CXRS 
  measurements~\cite{BellAIP2000} led to better agreement between
  experimental poloidal flows and NC expectations on JT-60U, MAST or 
  NSTX (see \cite{BellPoP2010}, and references therein), 
  in H-Mode plasmas or even in the presence of an Internal Transport
  Barrier (ITB).
  Moreover, recent measurements on Alcator C-Mod~\cite{KaganPPCF2011} 
  and ASDEX-Upgrade~\cite{ViezzerEPS2012} have been demonstrated 
  to follow NC theory in the absence of impurity density 
  asymmetries. Nevertheless, poloidal velocity measurements on JET 
  within an ITB~\cite{TalaNF2007}, and on DIII-D during H-mode and 
  quiescent H-mode~\cite{SolomonPoP2006}, are still not fully understood.
 
  On the other hand, toroidal flows in tokamaks are expected to be 
  dominated by mechanisms other than neoclassical, since toroidal 
  viscosity vanishes for axisymmetric systems. Nevertheless, the 
  presence of a sizeable toroidal ripple and/or non-axisymmetric magnetic 
  perturbations for ELM control in tokamak reactors could induce a 
  non-negligible toroidal viscosity. This demands experimental studies 
  on rotation in stellarator devices, since in such machines plasma cannot 
  rotate freely~\cite{HelanderPRL2008}, and the radial electric 
  field is determined by the ambipolarity condition on the 
  NC radial particle fluxes. Detailed comparisons of the NC radial 
  electric field with CXRS measurements have been performed on the 
  W7-AS stellarator, showing reasonable agreement 
  for most of the plasma scenarios studied~\cite{HirschPPCF2008}. 
  In addition, in the LHD and CHS stellarators, qualitative agreement 
  was found between radial electric field measurements 
  and calculations with simplified NC theories~\cite{IdaPoP1991,IdaPRL2001}. 
  The effect of the magnetic field ripple on the toroidal flows 
  was also investigated in CHS~\cite{IdaPoP1997} and 
  LHD~\cite{YoshinumaNF2009}, showing qualitative effects of helical and 
  toroidal ripple on spontaneous toroidal flow. Finally, a recent work in 
  HSX has shown a clear discrepancy between the experimental and NC radial 
  electric fields~\cite{BriesemeisterPPCF2012}. The authors claim that the 
  monoenergetic approximation used by DKES might be inadequate for radial electric fields close 
  to the helical resonance. Nonetheless, the parallel flow was well reproduced 
  after including momentum-correction techniques and impurities 
  in the calculations.
  
  In the TJ-II stellarator~\cite{SanchezNF2011}, the radial electric field 
  has been studied by means of the Heavy Ion Beam Probe (HIBP)~\cite{ChmygaEPS2002}
  and Doppler Reflectometry~\cite{EstradaPPCF2009} diagnostics. Measurements 
  from the former showed consistency with NC calculations in the electron root, 
  within a factor a two~\cite{ChmygaEPS2002}. However, the experimental values 
  systematically exceed the NC predictions in the ion-root 
  operation~\cite{DinklageIAEA2012}. In addition, passive spectroscopy 
  has been utilized to measured poloidal and toroidal rotations of impurities 
  \cite{ZurroFST2006,RapisardaEPS2005}. Qualitative agreement 
  was reported between the experimental poloidal flow and NC values, mainly 
  in the change of root as density increases~\cite{VelascoPRL2012}. In contrast, 
  significant discrepancies were found in the toroidal flows 
  for the different species studied and with NC calculations~\cite{RapisardaEPS2005}.

  In this work we use CXRS to measure fully ionized carbon impurity 
  flows in low density plasmas of the TJ-II stellarator and compare 
  them with neoclassical calculations. First, we present in
  section~\ref{sec:experimental_setup} the diagnostic technique, 
  together with a general treatment of the sightlines and flow geometry, 
  which allows the extraction of the two
  flux-surface averaged (FSA) flows from two velocity measurements. 
  These flux constants define an incompressible flow within a flux surface and 
  are related to the radial electric field and the bootstrap velocity. In section
  ~\ref{sec:incompressible_flows}, the 
  method is applied to two pairs from three measurements (poloidal, 
  outboard-toroidal and inboard-toroidal) to show that the spatial variation 
  of the flow is consistent with incompressibility. Finally, the measured 
  flows are compared with neoclassical predictions (Section~\ref{sec:comparison_NC}) 
  and good agreement is found for the radial electric field (in all the cases 
  studied) and bootstrap flows (in ECRH plasmas without momentum injection).
\section{Experimental set-up and data analysis}\label{sec:experimental_setup}
The TJ-II is a four-period heliac-type stellarator with major radius of $R_0=$1.5 m 
and averaged minor radius of $a\le$0.22 m~\cite{SanchezNF2011}. 
Hydrogen plasmas are created and heated by 2 gyrotrons operated at 
53.2 GHz (P$_{ECRH}\le$600 kW). During this heating phase central electron densities, 
n$_e(0)$, and temperatures, T$_e(0)$, up to $1.7\times 10^{19}$ m$^{-3}$ and $1$ keV, 
respectively, are achieved. In addition, plasmas can be maintained by 
two tangential Neutral Beam Injectors (NBI), each providing $\le$500 kW. 
As a result, plasmas with n$_e(0)\le5\times 10^{-19}$ m$^{-3}$ and T$_e(0)\le$400 eV 
are attained. The machine is operated with lithium wall coating, which allows 
density control, but results in 
a strong reduction in impurity concentrations and hence in impurity photon 
emission fluxes. In order to undertake CXRS measurements in such conditions, 
a high-throughput optical system and spectrograph, 
plus a high-efficiency detector, are needed~\cite{CarmonaRSI2006}. These are described next.

The CXRS process of interest in TJ-II involves electron capture from accelerated hydrogen 
by fully ionized carbon ions into a highly excited state  of C$^{5+}$, 
followed by spontaneous decay via photon emission, \emph{i.e.} the C VI line at 
529.07 nm ($n=8\rightarrow 7$). For this, a compact Diagnostic Neutral Beam Injector (DNBI) 
provides a 5 ms long pulse of neutral hydrogen accelerated to $30$ keV (the ratio of 
its full, half and third energy components is $90:8:2$)~\cite{CarmonaRSI2006}. 
Its $1/e$-radius at focus is $21$ mm.

For light collection, a bidirectional optical system, consisting of commercial 
camera lenses and 12-way fibre optic bundles, collects the C$^{5+}$ light 
across the outer half of the plasma diameter with $\sim 1$ cm spatial separation 
between lines of sight. These components are installed in top and bottom 
access ports located in the same machine sector as the DNBI. 
A second optical system with a 16-way fibre bundle, 
(12 equally spaced fibres along a row plus four additional fibres that straddle the row on both 
sides), views almost the complete beam/plasma interaction volume from a nearby 
tangential viewing access port. This provides $\sim3.4$ cm spatial separation between 
sightlines. Note: a retractable in-vacuum mirror located in the mouth of the access 
port is needed to redirect light to the lens as the port is shared with a neutral 
particle analyser diagnostic.

\begin{figure}[!h]
\centering
\includegraphics[width=7 cm]{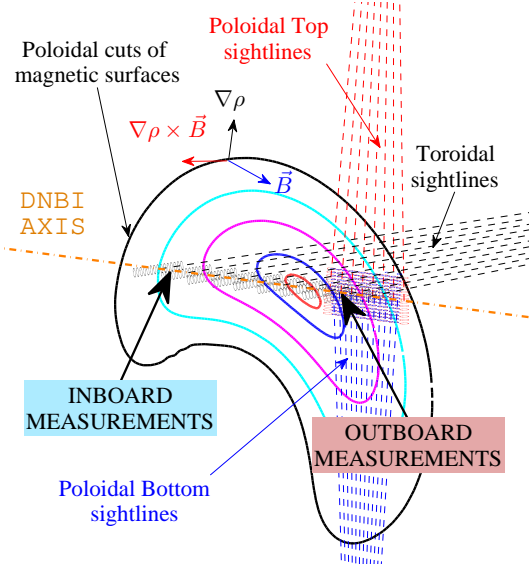}
\caption{\label{fig:sight-lines}Schematic diagram of CXRS diagnostic 
sightlines for the standard magnetic configuration. Only 8 Poloidal Top 
lines of sight are shown, as the outboard 4 do not provide useful data 
for this configuration. Inboard and outboard 
toroidal measurements are highlighted. In addition, poloidal cuts of 
magnetic surfaces are presented, together with the directions 
of concern for velocity measurements: radial ($\nabla\rho$), parallel 
(${\bf B}$) and ${\bf \nabla\rho}\times{\bf B}$. Note that none of these vectors is contained 
in the poloidal plane. The magnetic field vector, ${\bf B}$, points into the page.}
\end{figure} 

A schematic layout of the diagnostic sightlines is depicted in figure~\ref{fig:sight-lines}.
As observed, poloidal cuts of TJ-II magnetic surfaces are bean 
shaped and exhibit large flux compressions within a surface. 
The plasma minor radius region scanned by nearly symmetric poloidal views is
$\rho\in(0.25,0.85)$ in the standard magnetic configuration,
$100\_44\_64$ (where the nomenclature reflects currents in the central, helical and vertical 
field coils, respectively). Here, a normalised radius $\rho\equiv\sqrt{V/V_0}$ is defined, 
where $V$ and $V_0$ are the volumes enclosed by the surface of 
interest and the last closed magnetic surface, respectively. 
As explained in section~\ref{sec:data_analysis}, 
poloidal top fibres are used to correct for uncertainties in the 
C$^{5+}$ rest emission wavelength, and thus, do not provide additional 
information about the C$^{6+}$ distribution function. On the other hand,
the toroidal fibres cover both sides of the magnetic axis, from $\rho =-0.7$ to 
$\rho=0.6$ at 10 locations (2 of the 12 sightlines are obstructed by the vacuum chamber). 
The region in which both 
poloidal and toroidal measurements are taken is labelled as outboard, 
in analogy with tokamaks; while the zone where only toroidal 
measurements are made is labelled as inboard. 
The nomenclature $\rho\ge0$ and $\rho\le0$ is also utilized to define these regions. 
In the outboard region, poloidal and toroidal fibres 
view the same surfaces at $\rho\sim 0.2, 0.4$ and $0.6$. 
Therefore, the 2D-flow velocity is completely determined at these locations.
The redundant inboard-toroidal measurements will be used in 
section~\ref{sec:incompressible_flows} to verify whether the 
incompressibility assumption holds or not.

Light collected by all systems is transferred to the input of a light dispersion 
element, an holographic imaging spectrograph, modified with three $100~\mu$m-wide curved 
entrance slits (to compensate for the short focal length) and 
equipped with a transmission grating sandwiched between two BK7 prisms. It 
provides a focal-plane dispersion of $\sim 1.15$ nm/mm at $529$ nm~\cite{CarmonaRSI2006}. 
In addition, a narrow bandpass filter ($2.0\pm0.5$ nm at full-width at half-maximum) 
prevents spectral overlapping of the light from the three fibre arrays at the 
image plane. The set-up also includes a high-efficiency back-illuminated CCD camera 
and a fast mechanical shutter ($\ge 4.5$ ms time window). Using on-chip binning, 
multiple spectra are collected during discharges ($\le 300$ ms). 

\begin{figure}[!h]
\centering
\includegraphics[width=7 cm]{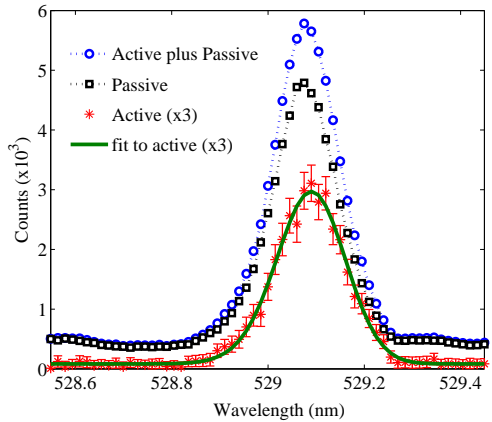}
\caption{\label{fig:raw_signals} DNBI-excited raw C-VI spectral line 
data (Active plus Passive) and
background emission (Passive), for two reproducible ECRH discharges. The active 
signal is obtained from the difference. A multi-Gaussian fit to the latter 
is presented.}
\end{figure}

In order to obtain neutral beam induced spectral line data, and hence spatially 
localized information, it is necessary to remove background -passive- C$^{5+}$ light from 
spectra. This is done by alternatively injecting and not injecting the DNBI 
into reproducible plasmas (shot-to-shot technique)~\cite{CarmonaRSI2006}). 
Then, by subtracting a background spectrum 
from the stimulated plus background -active plus passive- spectrum the CXRS data is obtained.
For ECRH plasmas, the active emission is a 
factor of $\sim5$ smaller than the passive one, see figure~\ref{fig:raw_signals}. 
In NBI heated plasmas, this ratio is reduced to a factor of $2$. A multi-Gaussian 
fit to the active signal is also shown,
and will be explained in Section~\ref{sec:data_analysis}.

Correctly performed instrument wavelength calibration and correction are essential 
if experimental uncertainties are to be minimized. Indeed, wavelength calibration 
is done regularly during TJ-II operation, as thermal variations in the instrument 
can be significant, \emph{e.g.}, $4$ $\mu$m K$^{-1}$ at the focal plane, this being 
equivalent to an offset of $\sim 2.6$ km s$^{-1}$ K$^{-1}$. For this, 
spectral lines from a neon pencil-type lamp are used to establish the wavelength 
dispersion at each fibre location on the focal plane~\cite{CarmonaRSI2006}. Finally, 
several well-studied effects are negligible for TJ-II or require correction before 
the Doppler shift of the beam induced C VI spectral lines can be determined. These 
effects, together with machine specific effects such as magnetic geometry and beam/sightline 
volumes, are handled by an in-house software package developed to analyse the data 
and obtain impurity ion temperature and velocities. 
These are described in more detail in Section \ref{sec:data_analysis}. 
\subsection{Optical alignment}
Alignment of the bidirectional poloidal sightlines is performed by 
illuminating each fibre bundle with an intense light source and observing, 
through a nearby view-port, the orientation and location of the resultant bright 
spots with respect to markings on the inside of the opposing vacuum flange. From 
these, the sightline paths through the neutral beam can be determined using a 
cross-sectional machine drawing and magnetic configuration maps. 

On the other hand, alignment of the tangential viewing sightlines 
involves aligning the in-vacuum mirror as well as orienting the fibre bundle ferrule. 
For this, a preliminary alignment is made by illuminating the fibres with the intense 
light source and observing, through a view-port located in a sector downstream of the 
neutral beam sector, the location of bright spots on the inside of the vacuum chamber. 
Then, after removing the light source, the fibre ends that attach to the spectrograph 
input fibre bundle, are connected to Avalanche Photodiodes (APD) whose bias voltages were 
previously adjusted to a common gain. Then, by repeatedly injecting the DNBI
into the TJ-II chamber, the mirror and fibre bundle are adjusted in-turn until all 
APD signals are maximized. 
Moreover, as an additional check, light from the sightlines that straddle the 12 principal 
sightlines are checked for symmetry about the beam centre. Finally, by repeating the 
fibre illumination procedure described before for the toroidal fibres, 
the location of bright spots on the vessel inboard wall are noted and cross-checked. 
Then, using a 3 dimensional computer aided design (CAD) software with the geometrical 
coordinates of the vacuum chamber, neutral beam and sightlines as inputs, the 
beam/sightline crossing points are found for the magnetic configuration of interest. 
\subsection{Carbon impurity temperature, density and velocity measurements}\label{sec:data_analysis}
\begin{figure}[!h]
\centering
\includegraphics[width=7 cm]{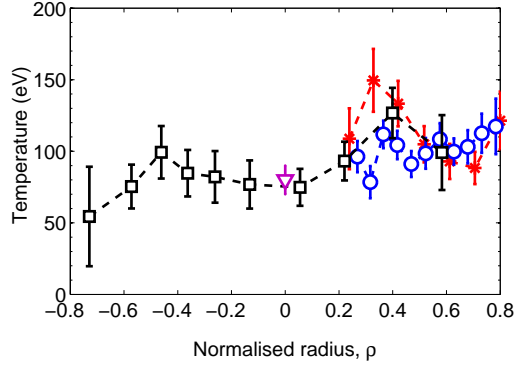}
\caption{\label{fig:ecrh_T_profile} Fully-ionized carbon ion temperature profile 
for a representative ECRH discharge (discharge $\#25801$, with 
$n_{{\rm e}}(0)\approx 0.5\times 10^{19}$ m$^{-3}$ 
and $T_{{\rm e}}(0)\approx 0.8$ keV), as measured
by poloidal top (blue circles), bottom (red asterisks) and toroidal sightlines 
(black squares). The central majority ion temperature measured by the NPA is shown as 
well (triangles).}
\end{figure}

In figure~\ref{fig:raw_signals}, a multi-Gaussian fit to an active signal obtained with 
the shot-to-shot technique is made using the expression:
\begin{equation}
I_{\rm C} = I_0\sum_ja_j\exp\left\{-\left(\frac{\lambda-\delta\lambda_j
-\lambda_{\rm c}}{\sigma}\right)^2\right\},
\end{equation}
where $a_j$ is the relative amplitude of each of the fine structure components 
of the C$^{5+}$ spectral line when
assuming a statistical population of the initial \emph{j}-levels, and normalized 
to the sum, $\sum_ja_j=1$. $\delta\lambda_j$ stands for the wavelength separation of the 
fine-structure components respect to the rest wavelength:
$\lambda_0=\sum_ja_j\lambda_j=5292.04~\mathring{\mathrm{A}}$ in vacuum~\cite{Hoang-BinhCompPhysCom2005}. 
In addition, $\lambda_{\rm c}$ measures the line centroid. Carbon temperatures are obtained 
from the parameter $\sigma$, after considering Zeeman broadening and a numerical 
deconvolution of the instrumental function~\cite{JArevaloHTPD2010}. 
A representative temperature profile of the ECRH phase is displayed in figure
\ref{fig:ecrh_T_profile}. It is found that CXRS temperature measurements from the three fibre 
arrays coincide, within error bars, on both sides of the magnetic axis. 
Indeed, the consistency of the temperature measurements from the three fibre arrays is 
used usually as an additional check of the diagnostic optical alignment.
For completeness, the majority ion temperature measured at $\rho\sim 0$ 
by the poloidal Neutral Particle Analyser (NPA), is shown.

Carbon impurity density measurements are more challenging, since
the beam density, $n_{{\rm b}}$, needs to be known. The latter can be roughly 
estimated by considering a simple pencil-like beam attenuation 
model~\cite{HutchinsonDiagnostics}, together with analytical expression to account 
for the beam profile in vacuum. Besides, the optical system is 
not absolutely calibrated at present.
However, an estimate can be made by assuming
that the relative efficiency of each fibre is proportional to the total intensity 
recorded during the instrument calibration procedure, \emph{i.e.},
\begin{equation}
\langle n_{{\rm C^{6+}}}\rangle_{\mathcal{V}_{{\rm los}}}\propto\frac{1}{\int_{\mathcal{V}_{{\rm los}}}
d^3{\bf r}n_{{\rm b}}}\frac{\int d\lambda I_\mathrm{C}}{\int d\lambda I_{\mathrm{Ne}}},
\end{equation}
where the integral of the beam atoms is performed in the sightlines measurement 
volume, $\mathcal{V}_{{\rm los}}$.

Finally, the velocity along sightlines, 
$v_l=\vec{v}\cdot\hat{e}_l$, is obtained from the line centroid shift, $\lambda_{\rm c}$:
\begin{equation}
\lambda_{\rm c}=(\lambda_0+\lambda_{\mathrm{cx}})\left(1+\frac{v_l}{c}\right)\approx
\lambda_0\left(1+\frac{v_l}{c}\right)+\lambda_{\mathrm{cx}}.
\end{equation}
Here, $\lambda_{\mathrm{cx}}$ accounts for possible deviations in the 
populations of the fine structure level from a complete statistical mixing,
that might cause a false velocity. This is removed by adding the 
redundant measured wavelengths from top and bottom poloidal sightlines, since 
they are almost symmetrically opposite. 

It is well documented that the daughter distribution function of C$^{5+}$ ions may exhibit 
some undesired velocities, usually named \emph{pseudo-velocities}, 
which arise from the energy dependence of the beam/carbon charge exchange cross-section 
as well as the gyromotion of the impurity during the de-excitation 
process~\cite{BellAIP2000}. In TJ-II, due to the relatively small magnetic 
field strength and temperatures, the only significant contribution comes from the 
former, which is orientated in the DNBI direction. These effects are only 
substantial for the toroidal view (since the poloidal sightlines view the beam almost 
perpendicularly), being of the order of $\sim 1-2$ km s$^{-1}$, at most. 
\subsection{Data Analysis}
The general form of the impurity velocity field is~\cite{HintonRMP1976}:
\begin{equation}
\label{eq:general_expression_flows}
{\bf u}=E(\rho)\frac{{\bf B}\times\nabla\rho}{B^2}+\left(\frac{\langle{\bf u}\cdot{\bf B}\rangle}
{\langle B^2\rangle}+E(\rho) h\right){\bf B},
\end{equation}
where the perpendicular velocity is given by ${\bf E}\times{\bf B}$ 
and diamagnetic contributions, and thus, 
\begin{equation}
E(\rho)=\frac{\rm{d}\Phi}{\rm{d}\rho}+\frac{1}{nZe}
\frac{\rm{d} p}{\rm{d}\rho}.
\end{equation}
Here, $n,(Ze)$ and $p$ are the density, charge and pressure of the impurity, respectively.
In addition, $\Phi$ is the electric potential and $\langle\cdot\rangle$ denotes flux-surface average. 
The last term in equation \eref{eq:general_expression_flows} is the well-know 
Pfirsch-Schl\"uter (PS) flow, which arises from the toroidicity of the magnetic field 
strength, and satisfies:
\begin{equation}
\label{eq:mag_eq_PS}
{\bf B}\cdot\nabla h = -{\bf B}\times\nabla \rho\cdot\nabla\left(\frac{1}{B^2}\right),
\quad \langle hB\rangle=0.
\end{equation}

In order to obtain flux-surface averaged (FSA) flows from poloidal and toroidal velocity measurements, it is 
convenient to define dimensionless vectors, which store the variation of 
the flow within the surface. These are:
\begin{eqnarray}
\label{eq:dimensionless_factors}
{\bf f}&=&-\frac{\langle B\rangle}{\langle|\nabla \rho|\rangle}
\left(\frac{{\bf B}\times\nabla\rho}{B^2}+h{\bf B}\right),\\
{\bf g}&=&\frac{\langle B\rangle}{\langle B^2\rangle}{\bf B}.
\end{eqnarray}
Note that the vector ${\bf f}$ has both perpendicular and parallel components, 
the latter coming from the PS contribution: 
${\bf f}\equiv {\bf f_\perp}+{\bf f_{{\rm PS}}}$. 
Then, impurity flows are expressed as:
\begin{equation}
{\bf u}={\bf f}U_\perp(\rho)+{\bf g}U_{\rm b}(\rho).
\end{equation}
Here, the following perpendicular and parallel FSA flows are defined:
\begin{eqnarray}
U_\perp(\rho)&\equiv&\frac{E_{\rm r}}{\langle B\rangle}-
\frac{1}{nZe}\frac{{\rm d}p}{{\rm d}\rho}
\frac{\langle |\nabla \rho|\rangle}{\langle B\rangle},\label{eq:Uperp}\\
U_{\rm b}(\rho)&\equiv&\frac{\langle{\bf u}\cdot{\bf B}\rangle}{\langle B\rangle},
\end{eqnarray}
where the FSA radial electric field is given by: 
$E_{\rm r}(\rho)\equiv -\langle |\nabla \rho|\rangle {\rm d\Phi}/{\rm d\rho}$.

Finally, taking the projections of the velocity vector ${\bf u}$
over the poloidal-bottom and toroidal sightlines ($u_{\rm p}={\bf u}\cdot\hat{e}_{{\rm p}}$ and
$u_{\rm t}={\bf u}\cdot\hat{e}_{{\rm t}}$, respectively) the FSA mean velocities are 
obtained from the measured flows:
\begin{equation}
\label{eq:fsa_flow_matrix}
{U_\perp(\rho) \choose U_{\rm b}(\rho)}=\frac{1}{\Delta}
\left(
\begin{array}{cc}
g_{\rm t} & -g_{\rm p} \\
-f_{\rm t} & f_{\rm p}
\end{array}
\right)
{u_{\rm p}\choose u_{\rm t}}.
\end{equation}
Here, $\Delta=g_{\rm t}f_{\rm p}-g_{\rm p}f_{\rm t}$, and the 
sub-index (p, t) 
denote the projection of a vector (${\bf u},{\bf g}$ or 
${\bf f}$) over the poloidal and toroidal lines of sight. Accordingly, 
the FSA flows can be obtained from two different measurements 
performed on a surface, but not necessarily at the same location. 
The results of this procedure are presented in the next section.
\begin{figure}[!h]
\centering
\includegraphics[width=7 cm]{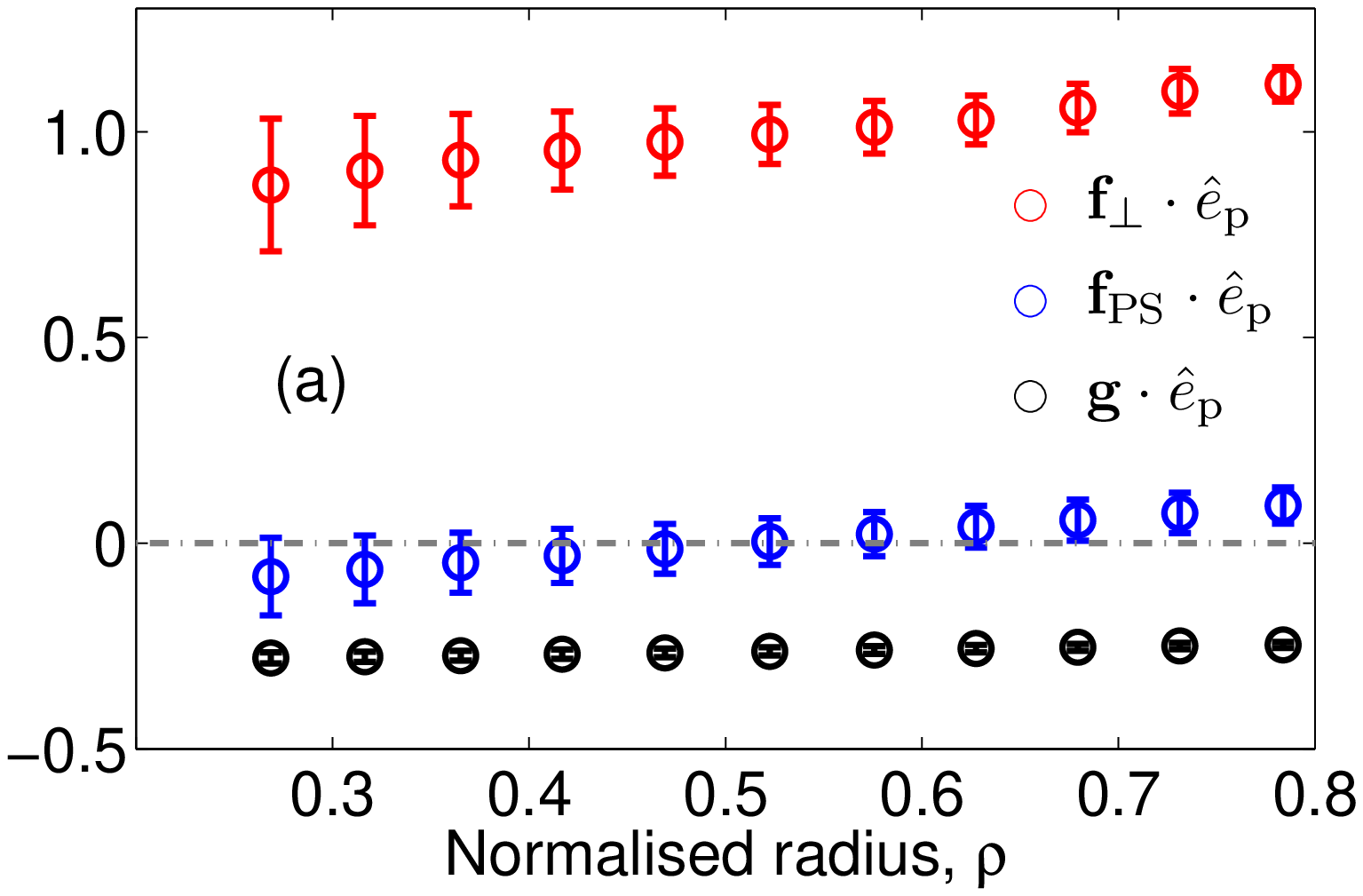}
\includegraphics[width=7 cm]{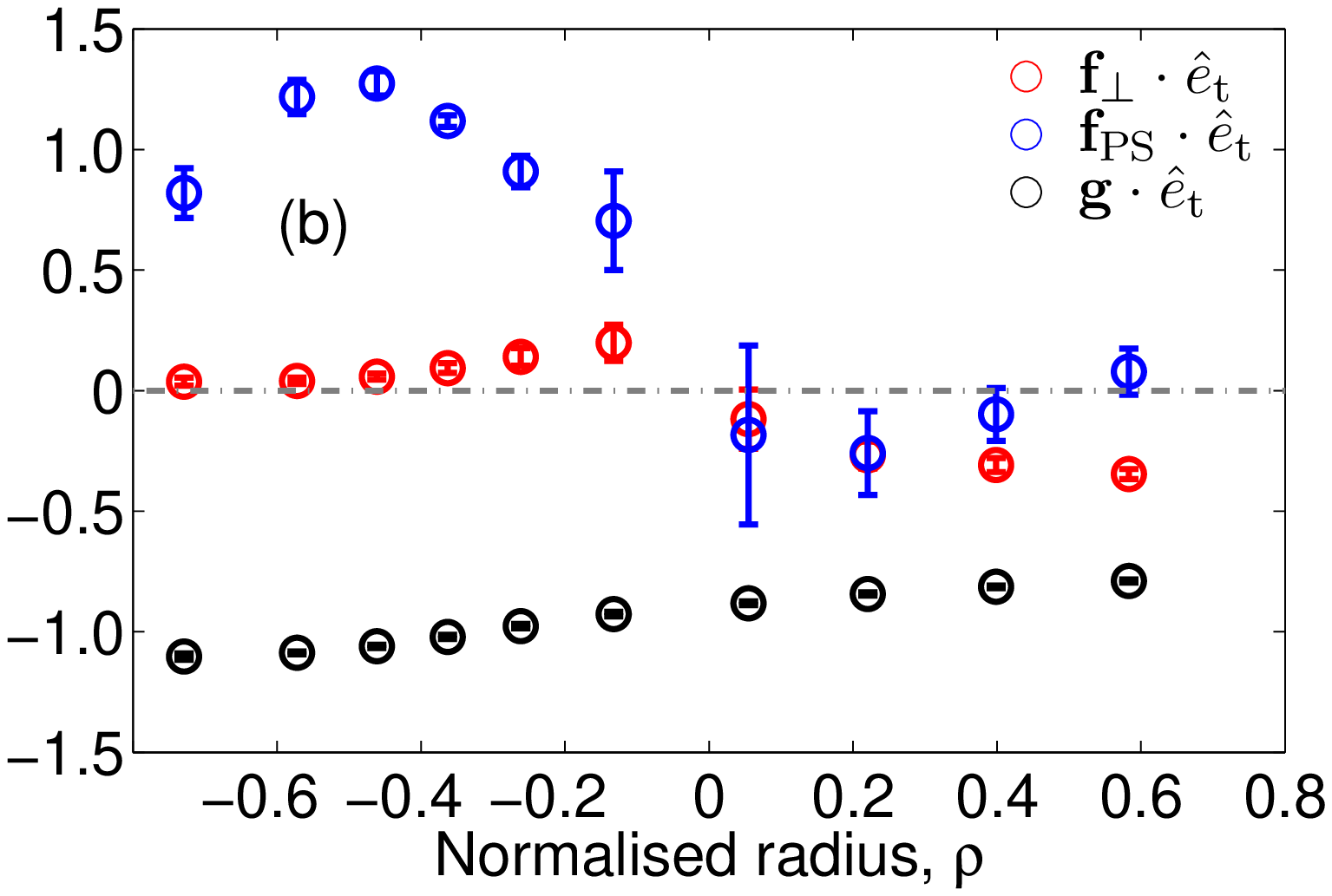}
\caption{\label{fig:geometrical_CXRS_factors} Dimensionless factors that store 
the variation of the flows on the surface for the (a) poloidal-bottom and 
(b) toroidal fibre arrays. The vector ${\bf f}$ has been 
split into their parallel and perpendicular components (${\bf f_{\rm PS}}={\bf f}\cdot\hat{b}$ 
and ${\bf f_\perp}=(\hat{b}\times{\bf f})\times\hat{b}$, respectively).}
\end{figure}

The projection of the geometrical vectors, ${\bf f}$ or ${\bf g}$, over 
the poloidal-bottom and toroidal sightlines is depicted in 
figure~\ref{fig:geometrical_CXRS_factors}. The vector ${\bf f}$ has been 
split into their parallel and perpendicular components, 
to illustrate their contributions on the velocity field. As expected, the poloidal 
velocity is dominated by the perpendicular contribution, although the parallel 
one (through the ${\bf g}\cdot\hat{e}_{{\rm p}}$ factor) has to be accounted for. 
The sample volumes located at the outboard are close to stagnation points of the PS 
flow and therefore its contribution to these measurements is modest, unlike in the 
inboard region where it can become dominant for toroidal velocities.
In addition, the mean values of a geometrical quantity, $X({\bf r})$, 
have been obtained through an average over the 
measurement volume for a given fibre, $\mathcal{V}_{{\rm los}}$:
\begin{equation}
\label{eq:CXRS_volume_average_II}
\langle X\rangle_{\mathcal{V}_{{\rm los}}}=
\frac{\int_{\mathcal{V}_{{\rm los}}}d^3{\bf r}n_{{\rm b}}n_{{\rm e}} X({\bf r})}
{\int_{\mathcal{V}_{{\rm los}}}d^3{\bf r}n_{{\rm b}}n_{{\rm e}}},
\end{equation}
Note that the integral should be weighted with the local carbon impurity density, 
rather than the electron density. Although the former is missing, we consider this 
approximation sufficiently good to obtain mean values.
The errors associated with geometrical quantities, $\varepsilon_{X}$, are given 
by its standard deviation in the measurement volume: 
$\varepsilon_{X}^2=\langle X^2\rangle_{\mathcal{V}_{{\rm los}}}-
\left(\langle X\rangle_{\mathcal{V}_{{\rm los}}}\right)^2$, and are propagated 
through the code. It can be observed that when moving towards the plasma centre, 
the dispersion in these factors increases, since the poloidal views become more perpendicular
to  the surfaces. The errors in the results will consist 
of the statistical ones, plus those due to the average in the measurement 
volume.
\section{Results}\label{sec:results}
\begin{figure}[!h]
\centering
\includegraphics[width=7cm]{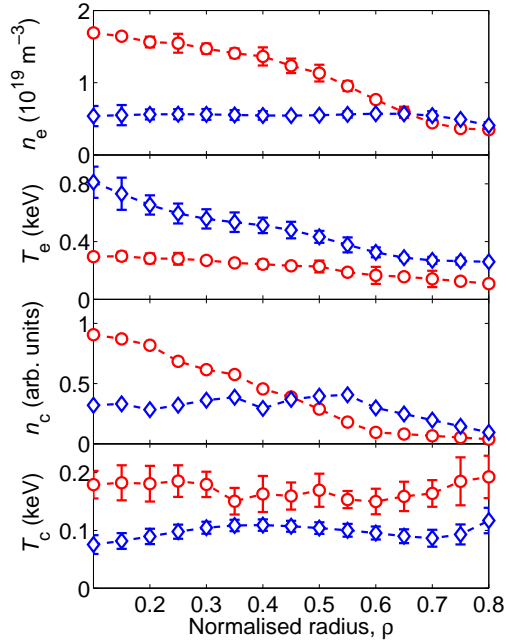}
\caption{\label{fig:plasma_profiles} Profiles of $n_{\rm e}$, $T_{\rm e}$, 
$n_{\rm c}$ and $T_{\rm c}$ (from top to bottom) for the ECRH scenario, 
shot $\#25801$ (blue diamonds); and the NBI case, discharge $\#28263$ (red circles).}
\end{figure}

In this work, two plasmas scenarios are considered, both being operated in the standard 
magnetic configuration. The first one, heated by ECRH,
is characterized by flat electron density ($n_{{\rm e}}(0)\approx 0.5\times 10^{19}$ m$^{-3}$) 
and peaked electron temperature profiles ($T_{{\rm e}}(0)\approx 0.8$ keV), see figure~\ref{fig:plasma_profiles}. 
However, ions remain cooler, $T_{{\rm i}}(0)\sim80$ eV 
($T_{{\rm i}}\approx T_c$ is assumed, see the comparison in figure~\ref{fig:ecrh_T_profile}),
with slightly hollow temperature and impurity density profiles. 
The data presented here corresponds to TJ-II 
discharge $\#25801$, which is representative of several similar discharges that are 
briefly discussed at the end of this section.

In the second scenario, plasmas are heated with a tangential NBI injected in the 
direction of the toroidal magnetic field. 
The resultant density profile is more peaked, with $n_{{\rm e}}(0)\approx 1.6
\times 10^{19}$ m$^{-3}$, whilst the electron temperature decreases to $T_{{\rm e}}(0)\sim 300$ eV. 
In contrast, the ion channel reaches temperatures up to $200$ eV. 
The discharge chosen for this case, $\#28263$, is not representative 
of fully developed NBI heated plasmas, as it corresponds to the first steps of the NBI phase, 
where the density grows rapidly. Nevertheless, it is included in this paper 
to highlight that for these low density plasmas the results are independent 
of the heating method.

Finally, the measurements presented are restricted to the region $|\rho|\le 0.8$, because CXRS 
measurements are not reliable at the edge of TJ-II, due to the poor statistics
and the possible presence of suprathermal ions~\cite{RapisardaPPCF2007}. 
\subsection{Flow incompressibility}\label{sec:incompressible_flows}
The assumption of an incompressible velocity field like the one in 
equation~\eref{eq:general_expression_flows} is widely used in 
present day CXRS data analysis. In this subsection we verify this 
by applying the method developed in the previous section, equation
\eref{eq:fsa_flow_matrix}, to the poloidal and toroidal velocities 
measured in the discharges presented in figure~\ref{fig:plasma_profiles}.

\begin{figure}[!h]
\centering
\includegraphics[width=7 cm]{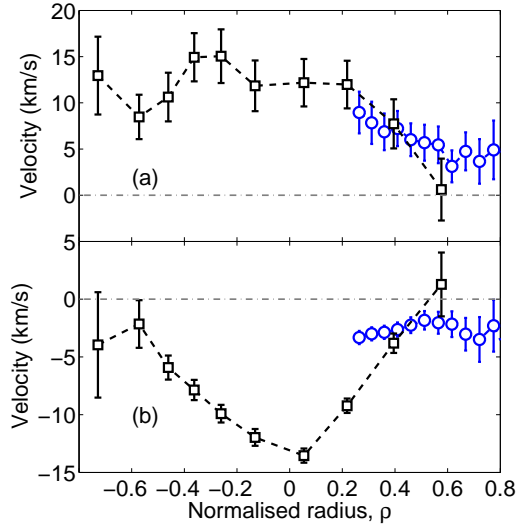}
\caption{\label{fig:raw_velocity}Carbon impurity ion flows observed by toroidal (black squares) 
and poloidal bottom (blue circles) sightlines, for the: (a)
ECRH plasma, discharge $\#25801$; and (b) NBI plasma, discharge $\#28263$.}
\end{figure} 
The observed velocity pattern, after correcting for fine structure effects and 
pseudo-velocities, is represented in figure~\ref{fig:raw_velocity} 
for the ECRH and NBI heating scenarios under study. Poloidal top velocity measurements 
are not shown, since they were already used to correct for fine structure effects, 
and so do not provide further information about poloidal flow.
In this figure, positive velocities corresponds to particles moving away from the observer.
In the ECRH plasma, the toroidal rotation is opposite to the magnetic field direction, 
while the toroidal flow is reversed in the NBI case. 
In addition, ECRH toroidal measurements exhibit a strong 
in/out asymmetry that is explained next.

\begin{figure}[!h]
\centering
\includegraphics[width=7 cm]{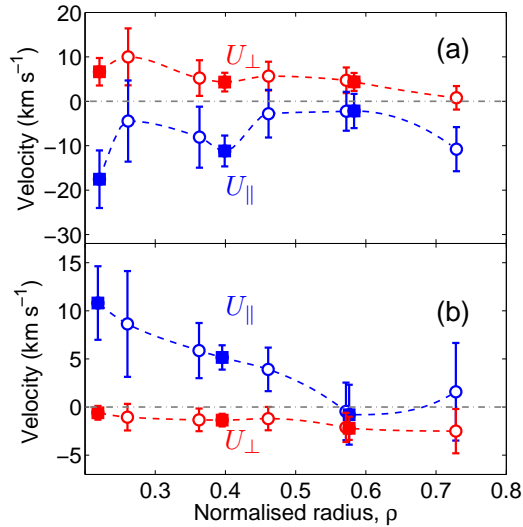}
\caption{\label{fig:fsa_flows} Flux-surface averaged flows $U_\perp(\rho)$ and 
$U_{\rm b}(\rho)$, for the: (a) 
ECRH discharge, $\#25801$; and (b) NBI plasma, $\#28263$. The circles 
correspond to toroidal measurements performed in the inboard region, and the 
squares to those taken in the outboard one.}
\end{figure}

As commented before, CXRS toroidal velocity measurements are performed on 
both sides of the magnetic axis in TJ-II. Therefore, they are suitable for the 
verification of flow incompressibility, due to its redundancy on several magnetic 
surfaces. In figure~\ref{fig:fsa_flows} the perpendicular, $U_\perp(\rho)$, and parallel, 
$U_{\rm b}(\rho)$, FSA flows obtained are plotted. The values that result from inboard toroidal 
measurements are presented as open circles, whilst the ones deduced from outboard 
toroidal velocities are squares. In both cases, the poloidal measurements inserted in 
equation \eref{eq:fsa_flow_matrix} come from the outboard region.

The agreement found between inboard/outboard measurements confirms the 
main hypothesis of the applied method, \emph{i.e.}, that impurity flows are incompressible
for the low density plasmas presented here. Indeed, to the authors' knowledge, this is the 
first time that this fundamental assumption of kinetic theory is checked 
with independent measurements at different locations of a surface (recent work 
at TCV and DIII-D~\cite{CamenemEPS2012,ChrystalRSI2012} extract the poloidal flow 
from toroidal in/out measurements, but are not compared with direct poloidal 
measurements, due to difficulties associated with such measurement in these 
tokamaks). 

The strong asymmetry in the toroidal velocity observed in the ECRH scenario, 
figure~\ref{fig:raw_velocity} (a), is explained by the larger -compared to 
the NBI case- perpendicular flow, which generates significant variations in 
the parallel velocity, through the local Pfirsch-Schl\"uter flow. As expected, 
there is a change from electron to ion root in the radial electric field with 
increasing density~\cite{VelascoPRL2012}, and therefore, a change of sign 
in $U_\perp$ (the impurity diamagnetic contribution is negligible, 
except for $\rho\sim 0.7$ in the NBI case, where it reaches $\sim 1$ km s$^{-1}$). 
Finally, a positive parallel velocity is observed in the NBI plasmas, which 
is compatible with the injection of the beam in the direction of 
the magnetic field. This will be discussed in the next subsection.

As an indication of the importance of this previous check on the flow 
spatial variation prior to its comparison with neoclassical estimates,  
we note that in higher density NBI plasmas a \emph{compressible} flow 
variation is observed in TJ-II. This makes the comparison of parallel 
velocities with standard neoclassical theory not straightforward. 
Indeed, impurity density variations within a flux surface have 
been theoretically predicted in the presence of large ion pressure 
gradients \cite{Helander_PoP1998}, and recently observed in several 
tokamak devices (Alcator C-Mod~\cite{MarrPPCF2012} and 
ASDEX-Upgrade~\cite{PuterichNF2012}). The asymmetry is related to 
ion-impurity friction and ultimately to impurity radial fluxes, 
which makes its comprehension particularly relevant for nuclear 
fusion. The analysis of such plasmas will be presented in a subsequent article.
\subsection{Comparison with Neoclassical theory}\label{sec:comparison_NC}
  
The Drift Kinetic Equation is solved by using the numerical code 
DKES~\cite{hirshman1986dkes} complemented with momentum correction 
techniques. DKES allows to calculate the 
monoenergetic transport coefficients of the magnetic 
configuration. These coefficients are convoluted with a 
Maxwellian distribution function which in turn depends on local 
density and temperature. Inclusion of the thermodynamical forces 
(gradients of density and temperature and radial electric field) 
allows the FSA neoclassical fluxes to be estimated. The 
neoclassical radial electric field itself is found iteratively by 
imposing ambipolarity on the particle radial fluxes. Specific 
details of the procedure are found 
in~\cite{VelascoPPCF2011,VelascoPPCF2012}.

The inputs for this calculation are the magnetic 
field equilibrium and TJ-II density and temperature profiles. 
The former is taken to be the vacuum equilibrium, 
for the low $\beta$ plasmas studied here. The latter require integration of data 
from the Thomson Scattering, Helium beam probe, Reflectometry, 
Interferometry (see~\cite{milligen2011bayes} and references there in) and 
CXRS diagnostics. The uncertainties in the transport coefficients and in the 
measured profiles are propagated to the final theoretical estimate, as 
in~\cite{VelascoPPCF2011}.

  In order to estimate accurately the parallel and poloidal momentum
  balance, one should make the calculation by including protons,
  electrons and main impurities present in the plasma.
  However, for the lithium-coated
  plasmas studied in this work, the impurity concentration is low
  \cite{SanchezNF2011} and does not modify the poloidal
  momentum balance, leaving $E_{\rm r}$ unchanged. Since the impurity
  diamagnetic term is small, this completely determines the poloidal
  impurity rotation. On the other hand, once one ensures momentum
  conservation in the calculations (by including interspecies
  friction), the bootstrap flow of the trace impurities is shown to
  follow that of the bulk ions, which are calculated as in~\cite{VelascoPPCF2011}.
  Finally, 
  one may wonder if values of $E_{\rm r}$ close to resonances 
  could be found, as in \cite{BriesemeisterPPCF2012}. In such a case,
  the poloidal $E\times B$ and magnetic drifts 
  cancel out, producing a peaking of the radial fluxes, and 
  the local ansatz underlying DKES computations fails. 
  At TJ-II, this happens for very large values
  of $E_{\rm r}$~\cite{GuaspPPCF2000}. The agreement found in
  the present comparison supports the hypothesis in our
  calculation.

  Some general considerations can be made, following the results
  obtained for model plasma profiles~\cite{VelascoPPCF2011,VelascoPPCF2012}. 
  Low density ECRH plasmas in TJ-II are in the electron root:
  $E_{\rm r}$ is positive and large, and therefore it partially cancels the large
  contribution of the $T_{{\rm e}}$ gradient to the electron radial
  transport (see figure~\ref{fig:plasma_profiles}).
  This large radial electric field drives a large ion parallel flow,
  especially close to the centre, where the fraction of trapped
  particles is small and $E_{\rm r}$ is considerable. 
  On the contrary, NBI plasmas are in the ion root. In order to reduce
  the ion particle radial transport, driven by the ion temperature
  gradients, to the electron level, $E_{\rm r}$ is small and negative. 
  This partial cancellation of the ion
  channel has consequences in the parallel momentum balance as well,
  where the negative $E_{\rm r}$ leads to a small ion bootstrap current.

In the following, the results of a comparison between NC theory and CXRS 
measurements are presented. Since CXRS active data has been obtained from two different 
discharges (shot-to-shot technique), we compare the experimental data with 
the average of NC outcomes for both plasmas.

\begin{figure}[!h]
\centering
\includegraphics[width=7 cm]{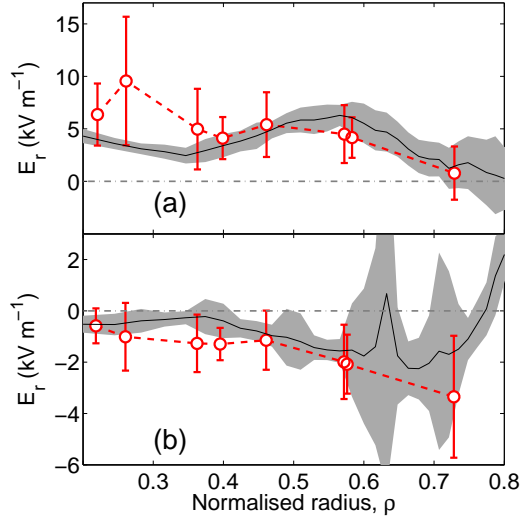}
\caption{\label{fig:Neo_Er} Comparison of Neoclassical (grey shading) and 
measured (red circles) $E_{\rm r}$, for the: (a) 
ECRH discharge, $\#25801$; and (b) NBI plasma, $\#28263$.}
\end{figure}
In figure~\ref{fig:Neo_Er}, the measured and calculated radial
electric field are shown. The experimental values 
are in agreement with previous results obtained in similar plasmas
with the HIBP diagnostic~\cite{ChmygaEPS2002}, and agree with the NC
predictions agree within the error bars. Nonetheless, the mean values
observed at the core of the ECRH plasma are somewhat larger
than that of the neoclassical estimate; however, the error bars are large. 
Moreover, the contribution to the experimental $E_{\rm r}$
from toroidal measurements,
$\propto -f_\parallel\hat{b}\cdot\hat{e}_{bot} u_t$, 
equation~\eref{eq:fsa_flow_matrix}, is similar to that of the 
poloidal flows, $\propto f_\parallel\hat{b}\cdot\hat{e}_{tor} u_p$, 
due to the high toroidal velocities observed in the core region,
see figures~\ref{fig:geometrical_CXRS_factors} and~\ref{fig:raw_velocity} (a).
This increases the uncertainties in $E_{\rm r}$, since toroidal sightlines 
are almost aligned with the magnetic field lines, and thus, the extraction 
of its perpendicular contribution is less reliable. 

In the NBI plasma, the experimental errors are reduced, see figure~\ref{fig:Neo_Er} (b).
The inferred radial electric field is 
in good quantitative agreement with neoclassical calculations. 
The error bars about the NC estimates in figure~\ref{fig:Neo_Er} 
correspond to dispersion in the ambipolar condition when one considers 
the not insignificant uncertainties in the plasma profiles, as in~\cite{VelascoPPCF2012}. 
In particular, for $\rho\ge0.6$, both electron and ion roots can result, as different 
values of density and temperature are considered within the error bars of 
figure~\ref{fig:plasma_profiles}, hence the large uncertainties.
Finally, the agreement found for this discharge in the radial electric field 
suggests that fast NBI ions do not contribute significantly to the radial 
ambipolar balance, although they inject parallel momentum, as discussed 
in the following text.

\begin{figure}[!h]
\centering
\includegraphics[width=7 cm]{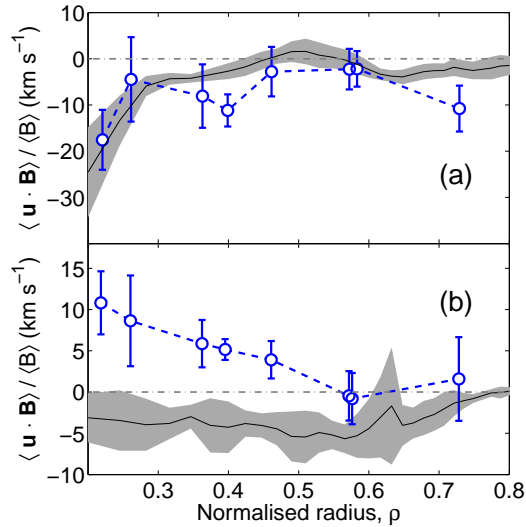}
\caption{\label{fig:Neo_boot} Comparison of Neoclassical and measured parallel flow, 
$U_{\rm b}$, for the: (a) 
ECRH discharge, $\#25801$; and (b) NBI plasma, $\#28263$.}
\end{figure}
Figure~\ref{fig:Neo_boot} (a) shows calculations of neoclassical bootstrap ion flow 
during the ECRH phase, together with the C$^{6+}$ parallel flux-surface flow, 
$\langle{\bf u}\cdot{\bf B}\rangle/\langle B\rangle$. 
A strong parallel rotation driven by the radial electric 
field is observed at the core region, that is well reproduced by the neoclassical predictions. 
In the higher density unbalanced NBI case, the
parallel flow consists of the neoclassical one (bootstrap), plus
the rotation induced by the injected neutrals. Indeed, these measurements
show that the latter dominates, being able to reverse sign. 
Since this is a low density NBI plasma, the injected fast neutrals can 
penetrate into the core. This would explain why the deviation of 
the parallel flow is maximum towards $\rho\sim0.2$, whilst measured 
parallel flow approaches bootstrap calculations for $\rho\ge0.6$.

\begin{figure}[!h]
\centering
\includegraphics[width=7 cm]{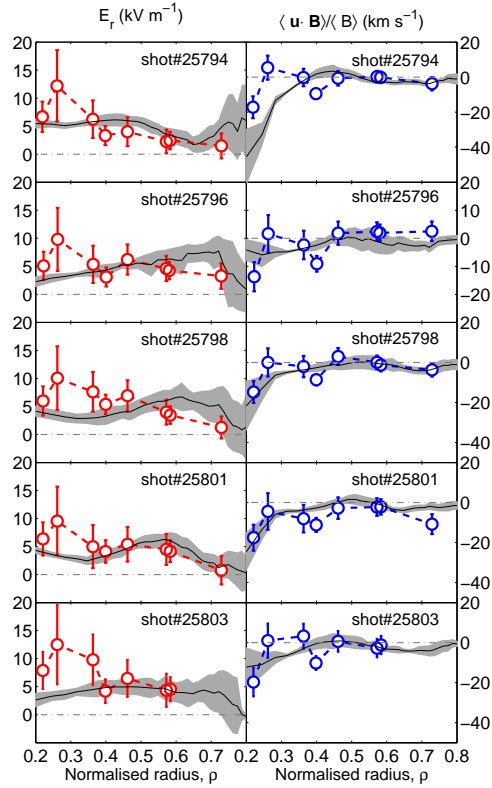}
\caption{\label{fig:Neo_ECRH_complete} Comparison of NC and measured radial 
electric field and bootstrap flow, for a 
set of ECRH discharges. Circles correspond
to CXRS measurements, and grey shading to NC estimates.}
\end{figure}

Finally, to highlight the reproducibility of the results in the ECRH scenario, 
the comparison of the measured radial electric field and bootstrap flow with 
NC calculations have been depicted in figure~\ref{fig:Neo_ECRH_complete}, for a 
set of similar ECRH discharges. The features observed in 
figures~\ref{fig:Neo_Er} (a) and \ref{fig:Neo_boot} (a) are replicated 
in these discharges, and the agreement with the NC calculations is repeated 
as well.
\section{Conclusion}
In summary, a comparison between CXRS carbon impurity flow measurements 
and NC calculations has been done. In addition, the basic form 
of the velocity field, \emph{i.e.} a divergence-free flow tangent
to magnetic surfaces, has been verified. To this end, a general treatment 
of sightlines and flow geometry has been applied to two pairs of three 
independent velocity measurements performed at different locations of 
a flux surface, and consistency with a 2D incompressible flow 
has been demonstrated. 

In addition, good quantitative agreement has been found between measured and NC radial electric fields for low density ECRH plasmas. Furthermore, a comparison of impurity parallel rotation with NC bootstrap flow has been done for the first time in TJ-II, showing consistency with the calculations in these plasmas. In an unbalanced NBI discharge with relatively low density, a similarly good agreement has been found for the radial electric field, while a change in the direction of parallel rotation (with respect to the NC calculations, which do not account for momentum injection) has been observed. This points to a minor role of the fast ion-driven radial current in the ambipolar balance (or poloidal momentum balance) of the TJ-II stellarator.

\section*{Acknowledgement}
The authors are indebted to the TJ-II experimental group.
They would like to thank B. Zurro for useful discussions, and
C. Hidalgo for his support. 
J. Ar\'evalo acknowledges financial support from the FPI grant awarded by CIEMAT 
(BOE resolution nº 171, 24/06/2008).

\end{document}